%% file: sample701.tex
\documentclass[twocolumn,trackchanges]{aastex701}
\usepackage{siunitx}
\usepackage{gensymb}
\usepackage{booktabs}
\usepackage{footnote}
\defcitealias{Borlaff2025}{B25}

\include{variables.tex}

\begin{document}

\title{SPHEREx confirms predictions for artificial satellite trail pollution in Low Earth Orbit}

\author[0000-0003-3249-4431]{Alejandro S. Borlaff}
\thanks{Corresponding author: \url{a.s.borlaff@nasa.gov}.\\ Project website: \url{https://sparkles.readthedocs.io/}}
\affiliation{NASA Ames Research Center, Moffett Field, CA 94035, USA}
\affiliation{Bay Area Environmental Research Institute, Moffett Field, California 94035, USA}
\affiliation{IAU Centre for the Protection of the Dark and Quiet Sky, 98-bis Blvd Arago, F–75014 Paris, France}
\email{a.s.borlaff@nasa.gov}

\author{Pamela M. Marcum}
\affiliation{NASA Ames Research Center, Moffett Field, CA 94035, USA}
\email{pamela.m.marcum@nasa.gov}

\author[0000-0002-8341-342X]{Steve B. Howell}
\affiliation{NASA Ames Research Center, Moffett Field, CA 94035, USA}
\email{steve.b.howell@nasa.gov}

\author[0000-0002-6278-9233]{Pablo M. Sánchez-Alarcón}
\affiliation{NASA Ames Research Center, Moffett Field, CA 94035, USA}
\email{pablo.m.sanchezalarcon@nasa.gov}

\author[0000-0001-6754-8933]{David Dubois}
\affiliation{NASA Ames Research Center, Moffett Field, CA 94035, USA}
\affiliation{Bay Area Environmental Research Institute, Moffett Field, California 94035, USA}
\affiliation{IAU Centre for the Protection of the Dark and Quiet Sky, 98-bis Blvd Arago, F–75014 Paris, France}
\affiliation{Centre de recherche sur les Ions, les Matériaux et la Photonique Université Caen Normandie, ENSICAEN, CNRS, CEA, Normandie Univ, CIMAP UMR6252, F-14000 CAEN}
\email{david.f.dubois@nasa.gov}

\author[0000-0002-7093-295X]{Jonathan McDowell}
\affiliation{Space Research Centre, Durham University, Stockton Rd, Durham, DH1 3LE, UK}
\affiliation{Planet4589 Research, 27 London Rd, Suite 6, Bromley BR1 1DG, UK}
\email{planet4589@gmail.com}

\begin{abstract}
The number of artificial satellites in Low Earth Orbit (LEO) is increasing at an exponential rate since 2019. Satellites are visible to both ground and space telescopes, and their bright emission in optical, infrared, and radio-wavelengths contaminate astronomical observations, degrading the data's scientific value. Recent simulations forecast that if all satellite constellations listed in current launch manifests are deployed to LEO, satellite trails will appear in up to 96\% of the images obtained by most space telescopes. In this article, we use the recently launched SPHEREx space telescope to corroborate these models. SPHEREx observations obtained between May and September 2025 indicate that $73.3^{+1.3}_{-1.2}\%$ of the images already show satellite trail contamination, with an average number of $N=2.18^{+0.11}_{-0.09}$ trails per exposure, providing observational validation of the published light contamination models. The observed satellite trails display highly inclined trajectories in agreement with the simulated ones. We discuss potential data reduction mitigation methods, and provide an updated satellite light pollution forecast for \emph{Hubble} and SPHEREx including the newer satellite constellations proposed in early 2026.   

\end{abstract}




\section{Introduction}\label{sec:intro}

Humanity has launched more satellites to Low Earth Orbit (LEO) in the last five years (2021 -- 2026) than in the previous seven decades of space flight combined\footnote{\texttt{Planet4589 Satellite Statistics: \url{https://planet4589.org/space/stats/active.html}}}. From 750 active satellites in 2000 to 15,000 satellites in early 2026, the exponential growth rate of LEO is driven by the rise of space telecommunication industry (i.e., \emph{Starlink, OneWeb, Guowang}). The unregulated use of space has prompted a rapid increase in the number of orbital debris \citep{ESA_Space_Report_2025}, orbital collision avoidance maneuvers \citep{pavanello+2025actaa236_770}, and even a potential depletion of the ozone layer driven by aluminum oxide (Al$_2$O$_3$) nanoparticles in the stratosphere from continuous payload reentries  \citep{ferreira+2024grl51_2024, maloney+2025na130_2024, revell+2025sci8_212}.

For astronomy, the side effects were immediate: scientists world-wide reported that reflected and self-emitted radiation from the satellites interfered with ground-based telescope operations at all wavelength ranges, from optical to radio \citep{mcdowell2020apj892_36, corbett+2020apj903_27,grigg+2023aap678_6, grigg+2025aap699_307}. Far from being immune from their orbital vantage point, even space telescopes like \emph{Hubble} \citep{kruk+2023nat7_262} and CHEOPS \citep{billot+2024arXiv2411.18326} started showing satellite trails on their images. 

Risk assessment studies for \emph{Vera C. Rubin} Observatory predict that 20\% of the midnight images will present satellite trails, and between 30\% to 80\% of all exposures obtained at the beginning and end of the night will be affected \citep{hainaut+2020aap636_121, tyson+2020aj160_226}. These assumptions were based on relatively conservative constellations of 26,000 -- 48,000 satellites. Those satellite population numbers roughly correspond to just the size of the \emph{Starlink} constellation. However, tens of different companies have already started planning and launching similar and larger constellations of their own \citep{mcdowell2020apj892_36}, a number that continues to grow. The total number of satellites currently (March 2026) proposed to the US Federal Communications Commission (FCC) and the International Telecommunication Union (ITU) is almost two million (1,800,000) satellites.

\begin{figure*}[htbp!]
 \begin{center}
\includegraphics[trim={0 0 0 0}, clip, width=\textwidth]{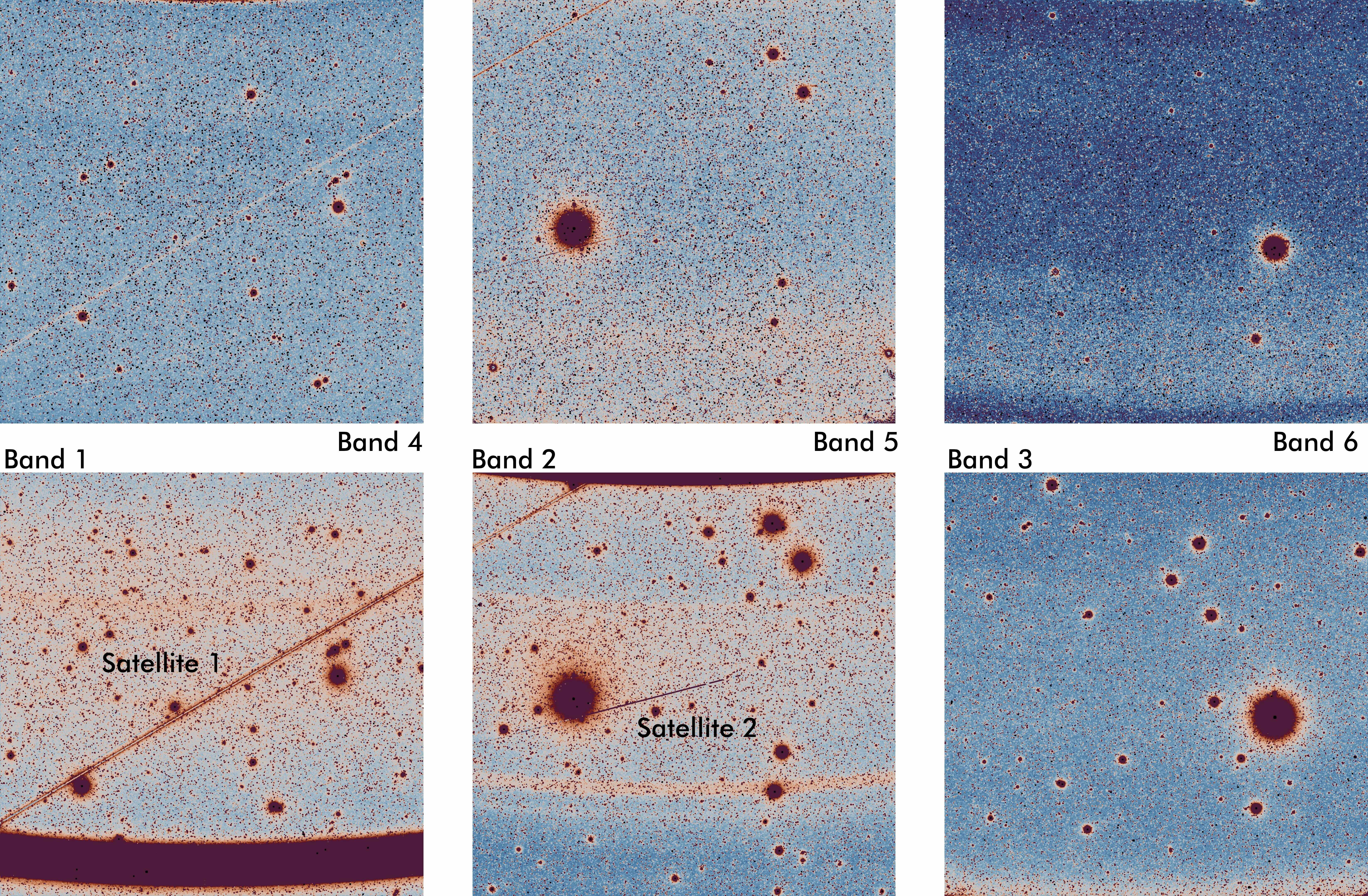}
\caption{Example of satellite trail identification on one of the SPHEREx exposures. Exposure ID: \texttt{2025W18$\_$1B$\_$0393$\_$3} (Obs. date: 2025-07-17 T22:12:03.217). From bottom to top, left to right: Band 1: ($\lambda=0.75-1.09\,\mu m$);
Band 2: ($\lambda=1.10-1.62\,\mu m$); 
Band 3: ($\lambda=1.63-2.41\,\mu m$);
Band 4: ($\lambda=2.42-3.82\,\mu m$);
Band 5: ($\lambda=3.83-4.41\,\mu m$);
Band 6: ($\lambda=4.42-5.00\,\mu m$).} 
\label{fig:example_trails}
\end{center}
\end{figure*}

In a recent article (\citealt{Borlaff2025}; hereafter \citetalias{Borlaff2025}), we forecast the impact of artificial satellite light contamination on a series of space telescopes including \emph{Hubble} Space Telescope and the Spectro-Photometer for the History of the Universe, Epoch of Reionization, and Ices Explorer \citep[SPHEREx,][]{feder+2024apj972_68, bock+2026apj999_139}. Our work estimates the number of satellite trails per exposure as a function of the increasing number of satellite telecommunication constellations, starting from 1000 to 1,000,000 satellites. The results show that if the proposed constellations are completed (560,000 satellites), one third of \emph{Hubble}'s images will be contaminated, while SPHEREx will have $>96\%$ of their exposures affected, with $5.6^{+0.3}_{-0.3}$ trails per exposure. Artificial light contamination from satellites is clearly a growing threat for ground and space-telescopes. 

On March 11, 2025 SPHEREx was launched, starting science operations on May 1st, 2025. The models from \citetalias{Borlaff2025} predict that SPHEREx should detect an average of $2.42^{+0.28}_{-0.28}$ satellite trails per exposure (over the complete 6 detectors covering $11\degree\times3.5\degree$) in 2025, assuming a total number of 20,000 satellites in orbit. The main purpose of this paper is to corroborate those claims, measuring the number of satellite trails in real SPHEREx science images obtained during the first year of operations (2025) and confirm or disprove the artificial satellite trail pollution forecast presented in \citetalias{Borlaff2025}. This paper is organized as follows. The datasets and methodology pipeline is described in Sec.\,\ref{sec:methods}. The results are presented in Sec.\,\ref{sec:results}. The discussion and conclusions are presented in Sec.\,\ref{sec:dis_and_con}.

\section{Methods}\label{sec:methods}

To determine the number of satellite trails in SPHEREx observations, we analyzed 6000 images from the IRSA/IPAC SPHEREx Archive\footnote{SPHEREx Archive: \url{https://irsa.ipac.caltech.edu/applications/spherex/}} taken at random pointings in the sky. SPHEREx orbits on a terminator-aligned sun-synchronous trajectory at 700 km over the surface of the Earth. The telescope has a $11\degree\times3.5\degree$ field of view, with a focal plane split by a dichroic beam into two focal plane assemblies (FPAs) that image the sky simultaneously with $2\times3$ HgCdTe H2RG detectors. The six detectors deliver wavelength coverage in six independent bands split between $\lambda=0.75-5.00\,\mu$m. Automatic satellite detection algorithms are still a work-in-progress task for SPHEREx images (see Sec.\,\ref{sec:limitations}). Consequentially, the satellite trail identification was performed visually using level 2 (calibrated) images, over all detectors simultaneously (see Fig.\,\ref{fig:example_trails}). This method allows the identification of satellite trails crossing multiple detectors on the same exposure or appearing on different bands, preventing duplicated identifications. Elongated structures completely aligned with the vertical or horizontal pixel grid were automatically discarded, to prevent false positives from charge transfer inefficiency and other readout systematics. We consider the limitations of visual identification in Sec.\,\ref{sec:limitations}.

Importantly, SPHEREx H2RG detector arrays use an on-board sample up-the-ramp (SUR) algorithm to measure fluxes on each pixel \citep[][]{zemcov+2016na5_1650007,crill+2020inproceedings_114430I}. Under this method, multiple non-destructive read-outs are performed every $\sim1.5$\,s during a total exposure time of 112\,s ($N_{\rm{reads}}\sim73$ readouts). Charge is accumulated over time on each pixel, providing a measure of the photon rate during the exposure. This method enables on-board flagging of charge overflow or saturation and transient conditions (usually cosmic rays, but also moving satellites), triggered by sudden changes in the photon rate. If a transient is detected, the pipeline halts the photon accumulation for the affected pixel. 

While the SUR algorithm reduces the impact of cosmic rays, it also  distorts the measured photon flux from satellite trails. 
Since SUR is calibrated for cosmic rays, pixel-gating is triggered only on bright satellite trails. For most satellites, SUR only masks the core of the satellite trail, leaving significant residuals on the wings of the point spread function. As a consequence, satellite trails remain detectable on SPHEREx calibrated single exposures through their residuals on the image, allowing their visual identification. However, we emphasize that due to the SUR algorithm 1) some trails might be undetectable after processing (false negative) and 2) photometric information, such as surface brightness magnitude, is not measurable from the available level 2 frames. Those limitations do not compromise the purpose of this work, which is to compare the observed satellite trail rate per exposure with the simulated contamination models. 

\begin{figure*}[tbp!]
 \begin{center}
\includegraphics[trim={0 0 0 0}, clip, width=\textwidth]{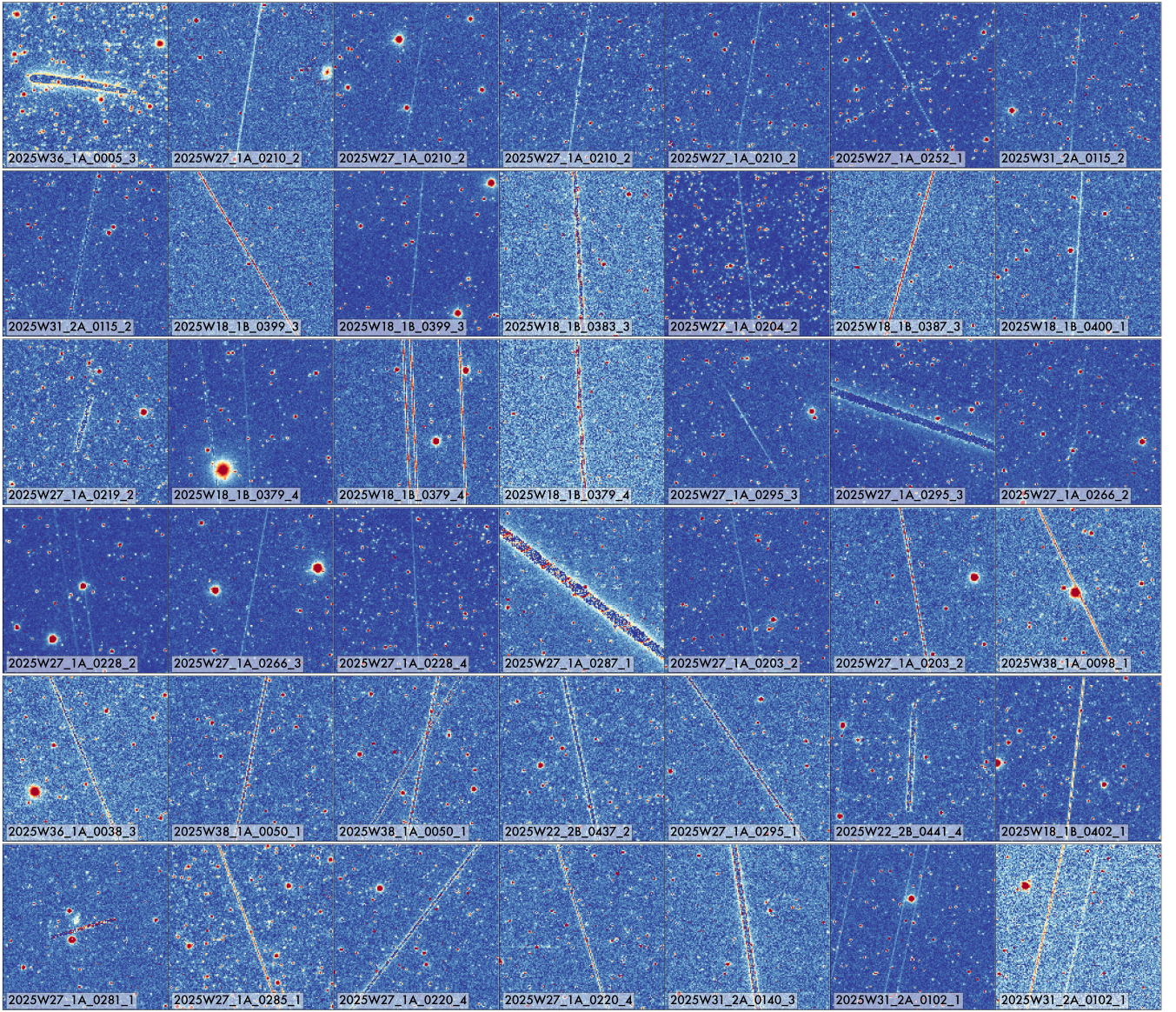}
\caption{Examples of the different morphologies of artificial satellite trails in SPHEREx science images. Brighter events tend to display broader trails with low signal residuals in their cores due to the effect of sample up-the-ramp (SUR) algorithm. Labels on each panel display the observation ID.} 
\label{fig:examples_trails}
\end{center}
\end{figure*}

One example of the satellite trail identification is presented on Fig.\,\ref{fig:example_trails}. In this exposure, the six detectors are inspected simultaneously. The lower (Bands 1, 2, 3) and upper (Bands 4, 5, 6) each scan the same $11\degree\times3.5\degree$ region of the sky ($\alpha$, $\delta$ = 342.7, -57.8) on different spectral bands. Bands 1, 2, 4, and 5 display a bright satellite trail (\#1) crossing between detectors. At the central detectors (Band 2) a second fainter satellite trail (\#2) is detected as well. The trail of satellite \#1 displays the characteristic impact of the SUR calibration algorithm, successfully masking the bright core of the trail but leaving significant ``railway-track" (Fig.\,\ref{fig:examples_trails}) residuals along the path of the satellite. Satellite 2, on the other hand, was not masked as a transient during readout. 

The identification of the satellite trails was performed by visual inspection of each exposure using a custom combination of \texttt{Python/Astropy} \citep{collaboration+2013aap558_33, collaboration+2018aj156_123, collaboration+2022apj935_167} and \texttt{SAODS9} \citep{softwareDS9}. For each exposure, the 6 bands are inspected and the satellite trails flagged as \texttt{SAODS9} line regions. Finally, the number, location, and orientation of the satellite trails was recorded. The process is repeated for a total of 6000 images (1.3\% of the archive at the same epochs) selected randomly over the first five months of the mission (May 2025 -- September 2025).

\section{Results}\label{sec:results} 
\subsection{Satellite trails in SPHEREx}\label{subsec:spherex_results}

\begin{figure*}[tbp!]
 \begin{center}
\includegraphics[trim={0 0 0 0}, clip, width=0.49\textwidth]{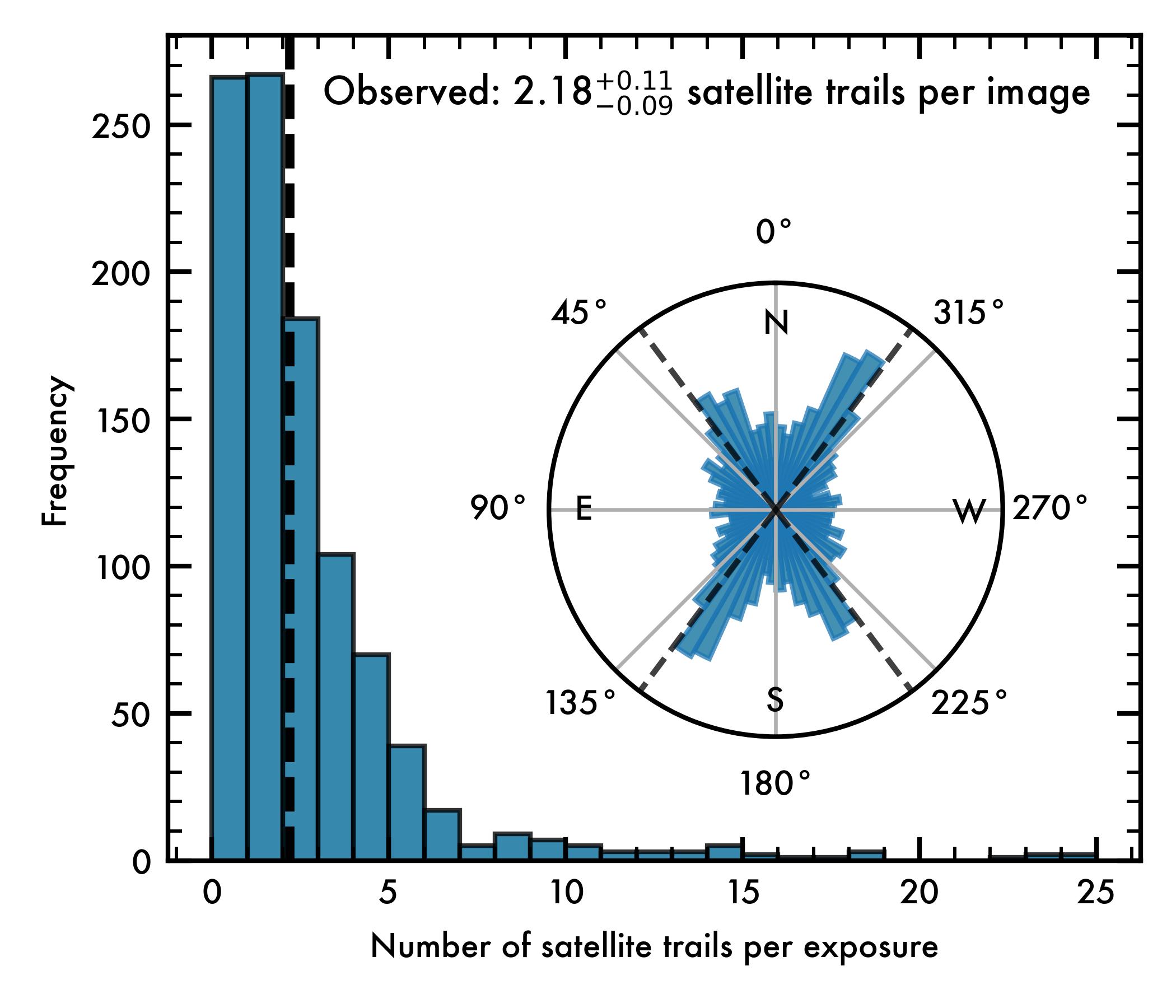}
\includegraphics[trim={0 0 0 0}, clip, width=0.49\textwidth]{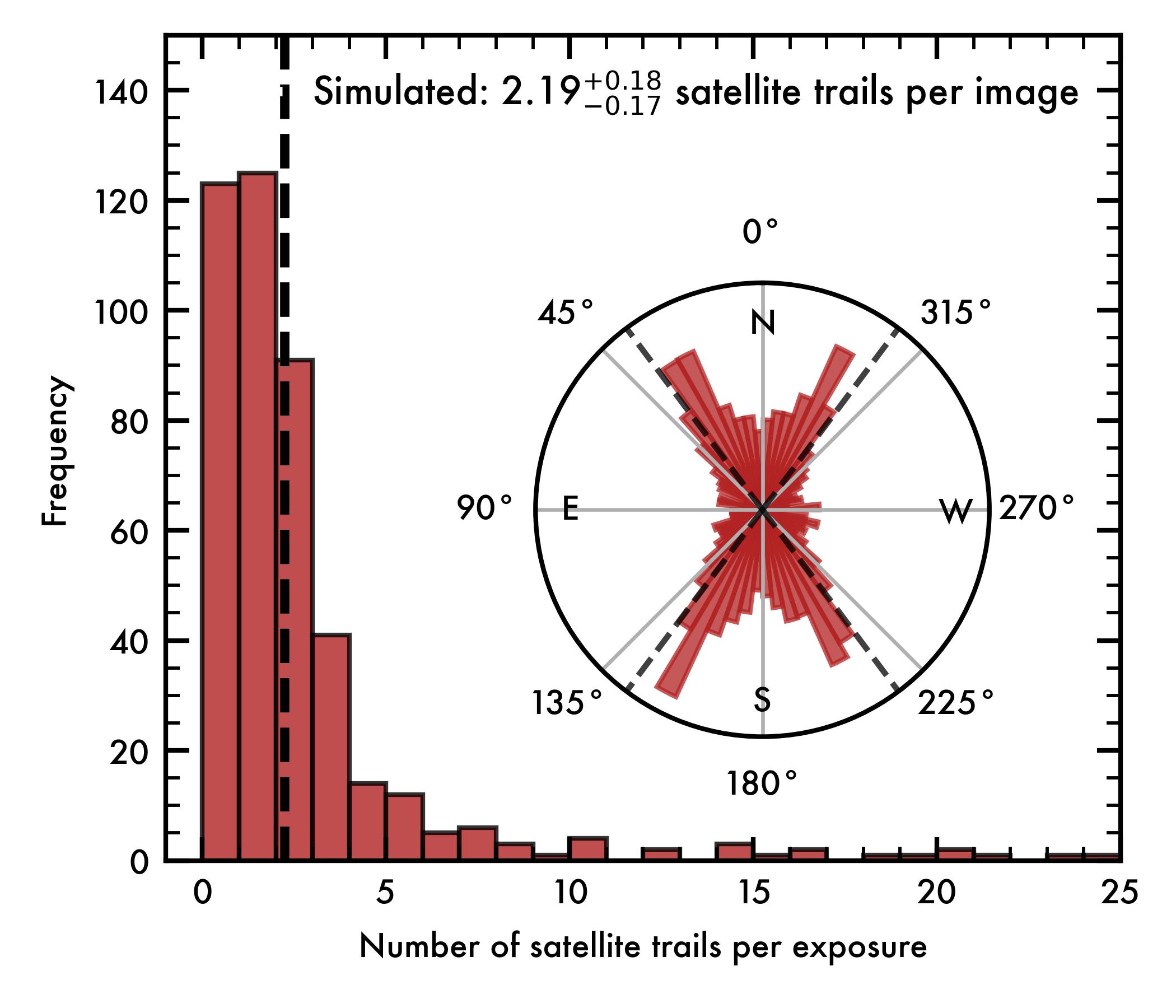}
\caption{Distribution of artificial satellite trails in SPHEREx science images. Left (blue) and right (red) panels represents the observations and simulated SPHEREx results. \emph{Main panel:} Histogram of number of trails per exposure. The average number of trails is $2.18^{+0.11}_{-0.09}$ in the observations, and $2.19^{+0.18}_{-0.17}$ trails per image in the simulations. \emph{Polar subpanel:} Histogram of satellite trail directions on equatorial (RA, Dec) coordinates. Note that the polar histogram has a 180$^{\circ}$ degeneracy since we do not know a priori the direction of the satellite along the trail. The diagonal dashed lines represent the average satellite orbit inclination (53.16$\degree$).} 
\label{fig:hist_trails}
\end{center}
\end{figure*}

The distribution of number of satellite trails per exposure is displayed on Fig.\,\ref{fig:hist_trails}. The results show a mean value of $N=2.18^{+0.11}_{-0.09}$ satellite trails per SPHEREx exposure in the analyzed sample. This number is compatible with the results from \citetalias{Borlaff2025} which predict an average number of $2.42^{+0.28}_{-0.28}$ trails per SPHEREx exposure with a total population of 20,000 satellites in LEO. In Sec.\,\ref{subsec:forecast_results} present an upgraded version of the \citetalias{Borlaff2025} models, including known satellite debris and sizes for objects with available information. In this updated model, the expected rate for SPHEREx is $2.19^{+0.18}_{-0.20}$ trails per exposure with 20,000 satellites in orbit. The observed satellite trail rate is then compared with the predicted contamination models, showing compatibility within $1\sigma$ confidence levels (see main panels in Fig.\,\ref{fig:hist_trails}). We use a two-sample non-parametric Kolmogorov-Smirnov test to analyze if both samples come from the same parent distribution. No statistical differences were found between the simulated and the observed trails ($p=0.40$).

The number of exposures with at least one trail is $73.3^{+1.3}_{-1.2}\%$. The observed rate of affected exposures also shows statistical compatibility with the results from the simulated models (see Sec.\,\ref{subsec:forecast_results}) which predict that $72.1\pm4.0\%$ of the SPHEREx images obtained in 2025 would be contaminated by at least one artificial satellite trail ($78.0 \pm 6.4\%$ in the \citetalias{Borlaff2025} original models). The original \citetalias{Borlaff2025} model predicts a higher (but 1$\sigma$ compatible) number of trails and fraction of affected exposures than the upgraded model. The difference is caused by the improved treatment of the cross-sectional sizes for the simulated satellites and debris. In the original version, all satellites have sizes between 1-125 m$^2$, while in the upgraded model, some objects (mostly debris) can have smaller sizes (down to 1 mm$^2$, see Sec.\,\ref{subsec:forecast_results}) and therefore, are less likely to be detected (see Fig.\,\ref{fig:appendix_mu_vs_size})

Notably, the observed satellite trails do not show a 
pure random orientation, they tend to be slightly diagonal from the North - South direction of the equatorial plane, displaying an X-shaped distribution projected in the sky. The histogram of trail orientations is presented in the subpanels in Fig.\,\ref{fig:hist_trails}. The distribution of satellite trail directions shows a peak in the number of trails at position angles around $\sim55 \degree$  from the East-West direction. This peculiar position angle distribution is also present on the simulated trails. We associate this distribution to the inclination of most existing satellites (the average inclination of the current satellite population is $i\sim53\pm27\degree$). This result suggests that the same population is responsible for the satellite trails in both observations and simulations.

\subsection{Satellite trail forecast}\label{subsec:forecast_results} 

\begin{figure*}[htbp!]
 \begin{center}
\includegraphics[trim={0 0 0 0}, clip, width=0.95\textwidth]{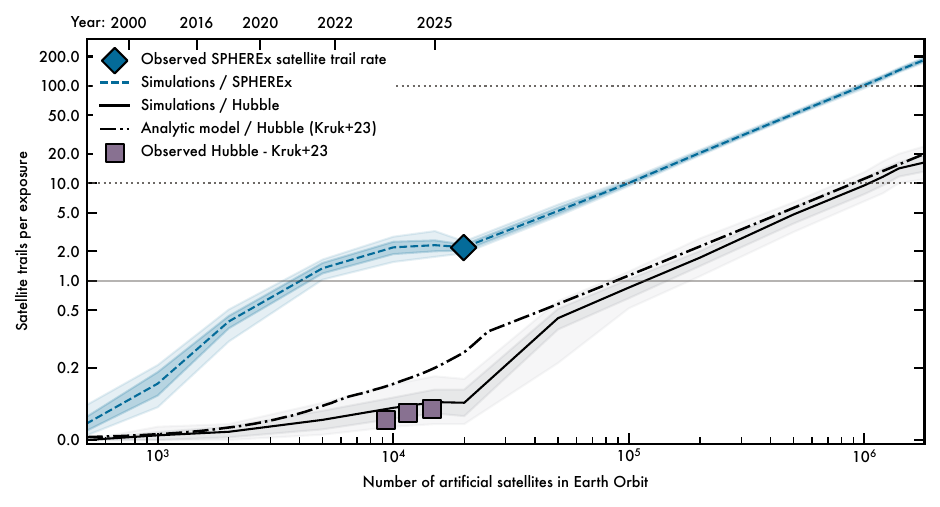}
\caption{Average number of satellite trails per exposure observed in SPHEREx observations compared against the predicted trail rate from \citetalias{Borlaff2025}, as a function of the population of artificial satellites in Earth orbit (lower $x-$axis) and epoch (upper $x-$axis). \emph{Blue diamond}: Observed average number of satellite trails in SPHEREx images. The different lines represent simulated models for the following observatories. \emph{Blue:} SPHEREx. \emph{Black:} \emph{Hubble} Space Telescope. Contours represent the 95\% confidence levels for the mean number of trails. \emph{Dashed-dotted line:} Predicted number of satellite trails based on \citet{kruk+2023nat7_262}. \emph{Grey squares:} Observer trails in \emph{Hubble} from 2002--2021 \citep{kruk+2023nat7_262}. \emph{Horizontal solid line:} One trail per exposure critical contamination level.} 
\label{fig:trails_vs_time}
\end{center}
\end{figure*}

Figure \ref{fig:trails_vs_time} compares the observed number of satellite trails per exposure with the predictions formulated in \citetalias{Borlaff2025}. The models are based on the same methodology as the original paper, including three major upgrades:

\begin{enumerate}

    \item Satellite debris with available cross-sectional area estimations (\texttt{Planet4589}\footnote{\texttt{Planet4589} Satellite filtering list: \url{https://planet4589.org/space/supporting/asb/asb.html}}) are now included in the models. We select the main bus diameter times span of longest appendage as the maximum cross-sectional area of the satellite in face-on orientation (\texttt{AREA 2}). Catalogued objects\footnote{Space-track: \url{https://www.space-track.org/}} with a) sizes smaller than $<1$ mm$^2$, b) not orbiting Earth, and c) flagged as \texttt{DOWN} (de-orbited), are removed from the simulation. 
    
    \item Satellites with known dimensions (i.e., SpaceX \emph{Starlink} Gen 1, 2, and xAI orbital data centers) have now a fixed area in the simulations. Those satellites for which their dimensions are unknown retain the original size distribution assumed in the original paper ($1$ -- $125$ m$^2$)

    \item The models include the latest FCC/ITU announcements (March 2026) from CTC1 and CTC2 (96,714 satellites each), and the SpaceX Orbital Data Centers (SXODC, 1,000,000 satellites). The total number of satellites considered (included existing debris, dead, and active satellites, and proposed constellations) adds up to 1,843,084 \footnote{\texttt{Planet4589}: Satellite Constellation list: \url{https://planet4589.org/space/con/conlist.html}}.
\end{enumerate}

At the satellite population levels corresponding to the analyzed SPHEREx images ($\sim$20,000 active and inactive satellites between May - September 2025), the observational satellite trail rate presented in Sec.\,\ref{subsec:spherex_results} is compatible within 1$\sigma$ with the predicted models. We did not find any significant seasonal variation during the analyzed period. The proposed orbits for the CTC1, CTC2, and SpaceX Orbital Data Centers include much higher orbits than the previous telecommunication constellations, resulting in a highly increased rate of satellite trails for all the considered telescopes. The results are summarized in Table \ref{tab:results_prop} on Appendix \ref{subsec:forecast_results}. If the proposed 1,840,000 satellites were deployed, SPHEREx would detect an average of $189^{+6.8}_{-5.6}$ trails per exposure (100\% of the images would show at least one satellite). For comparison, \emph{Hubble} would detect $16.6^{+3.0}_{-3.2}$ trails per image on average, $84.8\pm8.2\%$ of them showing at least one satellite.

We also compare the expected trail rate with the analytical model published by \citet{kruk+2023nat7_262} for \emph{Hubble} Space Telescope, to complement the verification of the simulation results. Integrating the probability of a satellite crossing the FOV of \emph{Hubble} with the altitude of each satellite layer allows to recover the expected number of satellite trails for the proposed satellite constellations (see their Eq. 1). The results from this independent model (see Fig.\,\ref{fig:trails_vs_time}) and the observations from \citet{kruk+2023nat7_262} show a very similar distribution of satellite trails as our model based on simulations, compatible within 1$\sigma$.

The results show a drastic increase of the expected number of satellite trails with respect to those reported in \citetalias{Borlaff2025} for SPHEREx ($101.5^{+3.6}_{-2.9}$ vs. $8.71^{+0.44}_{-0.44}$ assuming 1,000,000 satellites in orbit). The increase is caused by the inclusion of new proposed constellations, CTC1, CTC2 and Orbital Data Centers. These have much higher altitudes (1000 -- 21,500 km) than regular telecommunication satellites ($\lesssim$600 km) making them visible for the telescopes considered, including SPHEREx whose strict zenith angle prevent from observing satellites at lower altitudes. To compare, the simulations of \emph{Hubble} with orbital data centers present $\times$2--3 more trails than those without ($\times$10 for SPHEREx).

\section{Limitations}\label{sec:limitations} 

One major limitation of this work is the need for visual inspection to flag satellite trails on the SPHEREx exposures. Visual identification methods are notably less reproducible than automatic algorithms, and more time consuming. However, the implementation of satellite trail detection codes is still a field of research for SPHEREx images (JPL/SPHEREx pipeline team, priv. comm.). While automatic codes for detecting satellite trails have been developed for certain observatories like \emph{Hubble} Space Telescope \citep{borncamp+2016misc, stark+2022misc}, they are not immediately adaptable to other observatories with very different spectral ranges and detector properties (PSF, resolution, readout mode). In particular, as described in Sec.\,\ref{sec:methods}, SPHEREx SUR algorithm produces trails with "railway track" cores, not easily identifiable by algorithms designed for other telescopes. Alternatives for automatic satellite identification for the analyzed exposures in this work are an on-going effort by the SPHEREx team at JPL. One particular challenge for both automated and visual detection of satellite trails in SPHEREx is that most satellite trails do not show a constant brightness across the whole trail. This can be caused by several factors:

\begin{enumerate}
    \item The satellite had a different degree of illumination due to occultation with Earth's shadow.
    \item The satellite surfaces had significant change in orientation with respect of the telescope along the exposure.
    \item The satellite's albedo can have a significant variation with wavelength, causing SPHEREx detectors to be less sensitive to the object's emission on some regions of the detector. This is caused because the detector wavelength and bandwidth vary across the location on the SPHEREx detectors.
\end{enumerate}

To quantify the uncertainties associated with a visual detection of satellite trails, we compare the trail identifications from four different human observers. The measured uncertainty derived from the visual identification is $\pm0.43$ satellite trails, defined as the $1\sigma$ deviation between observers identifying the same image. This uncertainty is propagated across all average estimations in Sec.\,\ref{subsec:spherex_results}. In addition, we generated a series of synthetic trails on uncontaminated SPHEREx images. The synthetic trails were generated with different surface brightness magnitudes, ranging from 20 to 30 \magarc. The same team of observers identified the trails, showing a consistent visual limit in trail surface brightness of $\mu=24.69^{+0.77}_{-0.23}$ \magarc, compatible with the satellite trail magnitude limit estimated in \citet[][24.63 \magarc, $3\sigma$ detection]{Borlaff2025}.

Machine learning alternatives \citep{stoppa+2024aap692_199} depend highly on the type of datasets, in particular, frame shape, background noise, source properties, and spectral and photometric configuration of the telescope and detectors. While time consuming, visual identification of anomalies -- and combinations with citizen science efforts -- are very valuable, especially to serve as training samples for machine learning algorithms, as shown in \citet{kruk+2023nat7_262}.

The main objective of this article is to perform a first test of the average trail rate in SPHEREx, performing an evaluation of the observed morphologies and photometric properties of the events. The SPHEREx team maintains a database of anomalous images, \footnote{SPHEREx known anomalous images: \url{https://spherex.caltech.edu/page/anomalyimages}} which is complemented with the materials released with this article, providing a database of the identified satellite trails and exposure metadata in a Zenodo repository.

\section{Discussion and conclusions}\label{sec:dis_and_con} 

The results presented in this manuscript provide observational support to the models presented in \citetalias{Borlaff2025} on artificial satellite trail pollution, based on the first observations of SPHEREx. Two different statistics, the average number of satellite trails ($2.18^{+0.11}_{-0.09}$ per exposure) and the fraction of exposures contaminated by satellite trails ($73.3^{+1.3}_{-1.2}\%$) in SPHEREx images are statistically compatible at a 1$\sigma$ level with the predicted models at the 2025 satellite population in LEO (20,000 large debris, active, and inactive satellites). In addition, the orientation of the trails, measured as the position angle shows a similar distribution, caused by the inclination of the satellites, a feature predicted by simulations (see Fig.\,\ref{fig:hist_trails}). These findings provide observational confirmation to the prediction that tens to hundreds of satellite trails will appear on astronomical images of LEO space telescopes if the announced satellite constellations become operational. 

On a rapidly changing scene, more and more satellite operators are proposing new constellations with a potentially more harmful impact for the preservation of the dark and quiet skies. Far from trending to physically smaller dimensions, newer Direct-To-Cell telecommunication satellites (i.e., \emph{Starlink} Mini, 125 m$^2$, AST SpaceMobile, 220 m$^2$) are $4\times$ larger than older models (\emph{Starlink} V1.5, 26 m$^2$). \citetalias{Borlaff2025} assumed satellite sizes between 1 and 125 m$^2$. As other industries are stepping in, including AI servers \citep{Google_Suncatcher} and cloud data centers\footnote{\url{https://patents.justia.com/patent/20250108938}}\footnote{\url{https://www.thalesaleniaspace.com/en/news/ascend-new-alternative-terrestrial-datacenters}}, satellites with sizes up to $3600$ m$^2$ are being proposed. The massive increase in satellite size greatly cancels out the marginal benefits of the optical darkening attempts \citep[i.e., abandoned mitigation strategies like DarkSat or VisorSat,][]{tregloanreed+2021aap647_54, halferty+2022mnras516_1502, boley+2022aj163_199}. 

Optical reflections are just a fraction of the problem, since satellites also emit and reflect IR \citep{horiuchi+2020apj905_3}, and optical darkening might even increase their temperature, making them brighter in the IR. On the other side of the electromagnetic spectrum, satellite constellations are among the brightest objects for radio-telescopes like SKA \citep[up to 10$^6$ Jy per beam]{wayth+2022na8_011010, grigg+2023aap678_6} and LOFAR \citep{divruno+2023aap676_75, bassa+2024aap689_10}, not only due to the payload emissions and downlinks, but also reflecting terrestrial FM radio emissions \citep{grigg+2025aap699_307}. For SPHEREx and other multi-readout detector based telescopes (e.g., H2RG, H4RG), the capability to detect time-variation of the received flux from sources -- often applied to reject cosmic-rays -- suggests that satellites might be identifiable by using readout algorithms like SUR, enabling masking of some trails.

Remarkably, some private operators\footnote{Reflect Orbital AAS Assessment: \url{https://tinyurl.com/AASReflectOrbital}} have announced the intent of deploying 4000 purposely high-reflective satellites 324 m$^2$ in size aiming to offer ``sunlight on demand". If deployed, each satellite would reflect $4-5\times$ the brightness of the full Moon on $\sim5$ km diameter beam on Earth's surface, with a catastrophic impact to the quality of the dark sky and the Earth's inhabitants. Following the models of \citetalias{Borlaff2025}, we expand the surface brightness magnitude models of a trail caused by satellites ranging from 1 to 10,000 m$^2$ (see Fig.\,\ref{fig:appendix_mu_vs_size}). The results show that satellites as those proposed in 2025 by \emph{Google}, \emph{Sophia Space}, \emph{Thales Alenia} or \emph{Starcloud} would range between 21 and 17 \magarc\ in the optical range, and 17 and 12 \magarc\ in the IR, replacing the brightest sources in the night sky. Furthermore, the chemical impacts on the mesosphere and stratosphere composition and radiative balance resulting from reentry burns are largely unknown. Modeling efforts have only started to investigate the suspected role played by Al$_2$O$_3$-based aerosols on environmental chemistry and potential ozone depletion \citep{ferreira+2024grl51_2024, maloney+2025na130_2024}. Nucleation and photochemical processes of these aluminum oxide-based nanoparticles are thus likely to have broader impacts on Earth's mesospheric and stratospheric atmospheric chemistry.

Article 1 of UNESCO's Human rights for future generations declaration \citep{UNESCO_lalaguna} states that \emph{persons belonging to future generations have the right to an uncontaminated and undamaged Earth, including pure skies}. In a way similar to those that 
combat threats to human health from air, water, and soil pollution \citep{nihWaterPollution}, a growing need for strong regulations for the use of space are required now. The widespread contamination addressed in this paper affects multiple areas of astronomical research \citep{SATCON1, SATCON2}, such as the study of transient phenomena, deep imaging surveys, spectroscopic studies, and the search for Near-Earth Objects (NEOs), thereby hindering our ability to understand the Universe and protect the Earth \citep{ferreira+2024grl51_2024, maloney+2025na130_2024, revell+2025sci8_212}. The results presented in this article demonstrate that the satellite contamination forecasts are already taking place. 


\begin{acknowledgments}
The authors thank the anonymous referee for the careful reading and constructive suggestions that improved the manuscript significantly. This manuscript uses SPHEREx datasets from the SPHEREx Quick Release Spectral Images QR and QR2, made available by the Infrared Science Archive (IRSA) at IPAC, which is operated by the California Institute of Technology under contract with the National Aeronautics and Space Administration. The authors thank Brendan P. Crill and the SPHEREx team for their support and careful review. The authors acknowledge the support of the International Astronomical Union (IAU) Centre for the Protection of the Dark and Quiet Sky (CPS). The Centre coordinates collaborative and multidisciplinary international efforts from institutions and individuals working across multiple geographic areas, seeks to raise awareness, and mitigate the negative impact of satellite constellations on ground-based optical, infrared and radio astronomy observations as well as on humanity’s enjoyment of the night sky. Any opinions, findings, and conclusions or recommendations expressed in this material are those of the author(s) and do not necessarily reflect the views of the NASA, IAU, NSF NOIRLab, SKAO, ESO, or any host or member institution of the IAU CPS. The SPHEREx datasets are available through \citet{https://doi.org/10.26131/irsa629} and \citet{https://doi.org/10.26131/irsa652}.

A live version of the artificial satellite trail forecast is available at the project website\footnote{Project \texttt{SPARKLES}: \url{https://sparkles.readthedocs.io/en/latest/}}. The database containing the satellite trails identified in the analyzed SPHEREx images is publicly available in Zenodo, as an associated repository \doi{10.5281/zenodo.17957747}, with a \texttt{Jupyter-notebook} for reproduction of Figures \ref{fig:hist_trails} and \ref{fig:trails_vs_time}. 
\end{acknowledgments}

\begin{contribution}


\end{contribution}

%
\facilities{NASA/SPHEREx}

\software{\texttt{Astropy} \citep{collaboration+2013aap558_33, collaboration+2018aj156_123, collaboration+2022apj935_167}, \texttt{Skyfield} \citep{rhodes2019misc}}

\clearpage

\appendix
\section{Satellite trail surface brightness vs. cross sectional area}\label{appendix:mu_vs_size} 

\begin{figure*}[htbp!]
 \begin{center}
\includegraphics[trim={0 0 0 0}, clip, width=0.95\textwidth]{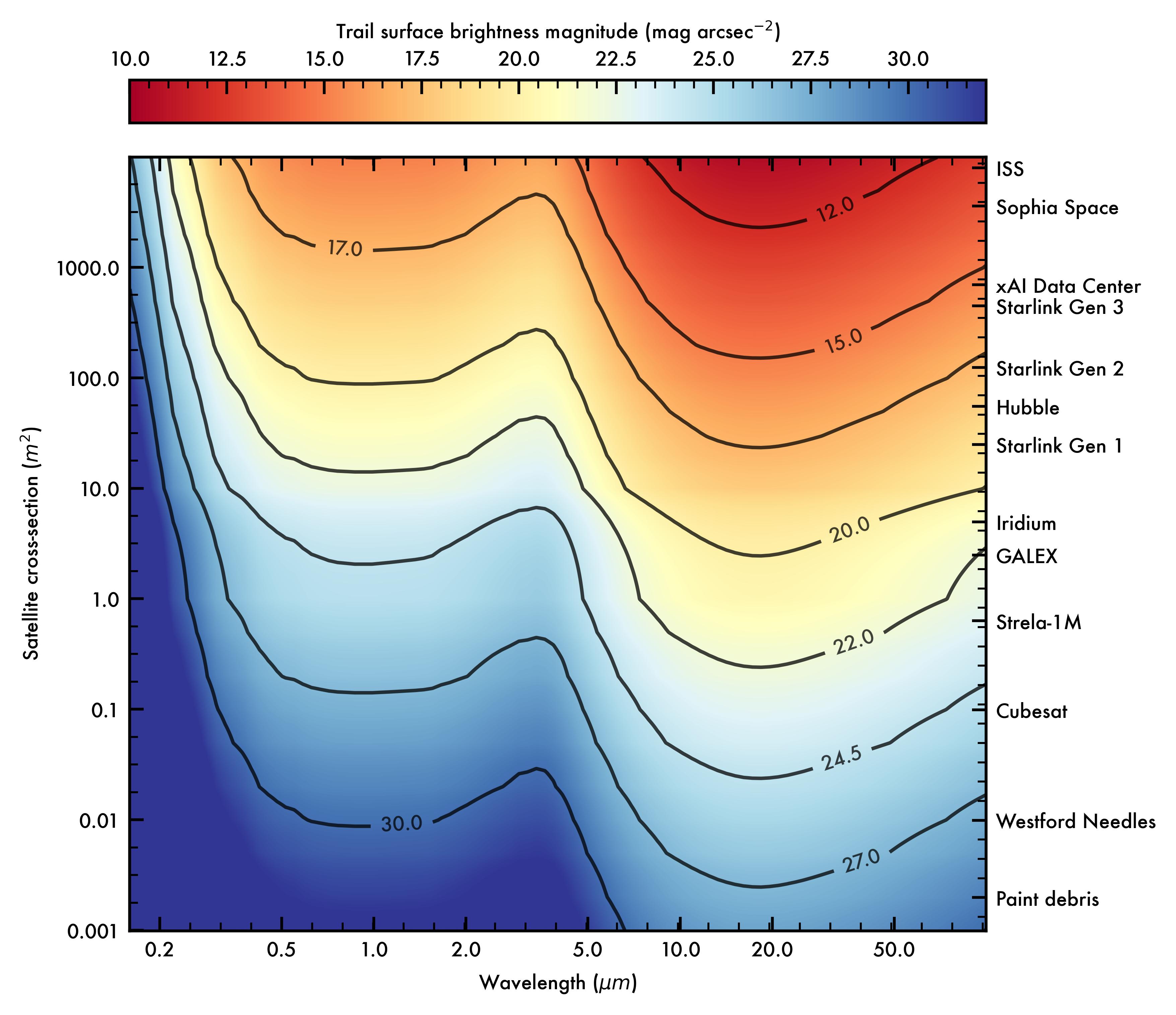}
\caption{Average satellite trail surface brightness spectral energy distribution as a function of the physical size of the satellites in Low Earth Orbit. Estimations obtained following the photometric model from \citetalias{Borlaff2025}, including Sunshine, Earthshine, Moonshine, thermal black-body radiation from the satellites ($T=280\pm3$K) and reflected thermal emission from the Earth ($T=290$ K).} 
\label{fig:appendix_mu_vs_size}
\end{center}
\end{figure*}

\clearpage

\section{Properties of the population of simulated satellite trails}\label{appendix:results_prop} 

\begin{table}[hbt!]
\caption{Properties of the observed population of satellite trails in the simulated exposures, as a function of the telescope and the number of satellites simulated ($N_{\rm sat} = $ 20,000 -- 1,840,000). Col.(1) Observable parameter. Col.(2) Telescope. Col.(3 -- 7) Estimated values as a function of the number of satellites. Group rows: a) Number of satellite trails per exposure. b) Exposures with at least one trail (\%). c) Area affected. d) Field of view covered by trails (\%).\label{tab:results_prop}}
\begin{tabular}{llccccc}
\toprule
 Observable & Telescope & \multicolumn{4}{c}{$N_{\rm{sat}}$: Number of satellites} \\
\midrule
 &  & 20,000 & 100,000 & 500,000 & 1,000,000 & 1,840,000 \\
(1) & (2) & (3) & (4) & (5) & (6) & (7)\\

\midrule
a) Trails per& \emph{Hubble} & $0.10^{+0.04}_{-0.04}$ & $0.85^{+0.18}_{-0.19}$ & $4.79^{+0.90}_{-0.93}$ & $9.5^{+1.8}_{-1.8}$ & $16.6^{+3.0}_{-3.2}$\\

  exposure  & \emph{SPHEREx}  & $2.19^{+0.18}_{-0.20}$ & $10.08^{+0.37}_{-0.36}$ & $51.4^{+1.5}_{-1.6}$ & $101.5^{+3.6}_{-2.9}$ & $189.1^{+6.7}_{-5.6}$\\
\midrule

b) Exposures with at & \emph{Hubble} & $8.1 \pm 2.4$ & $32.4 \pm 4.9$ & $57.4\pm6.5$ & $72.4\pm7.4$ & $84.8\pm8.2$\\

 least one trail (\%)  & \emph{SPHEREx} & $72.1\pm4.0$ & $100.0\pm8.0$ & $100.0\pm 8.0$ & $100.0\pm8.1$ & $100.0\pm8.3$\\

\midrule

c) Area affected & \emph{Hubble} ($\times10^{-2}$) & $0.22^{+0.08}_{-0.08}$ & $2.91^{+0.63}_{-0.65}$ & $26.6^{+5.0}_{-5.1}$ & $52.8^{+9.9}_{-9.9}$ & $104^{+19}_{-20}$\\

 (arcmin$^2$)    & SPHEREx ($\times10^{2}$)& $1.01^{+0.07}_{-0.08}$ & $10.8^{+0.40}_{-0.39}$ & $56.7^{+1.6}_{-1.8}$ & $112.0^{+4.0}_{-3.2}$ & $208.8^{+7.5}_{-6.2}$\\

\midrule

d) Field of view & \emph{Hubble} & $0.020^{+0.007}_{-0.007}$ & $0.260^{+0.056}_{-0.058}$ & $2.38^{+0.45}_{-0.46}$ & $4.73^{+0.89}_{-0.88}$ & $9.3^{+1.7}_{-1.8}$ \\

 affected  (\%) & \emph{SPHEREx}& $0.07^{+0.01}_{-0.01}$ & $0.76^{+0.03}_{-0.03}$ & $3.98^{+0.11}_{-0.13}$ & $7.88^{+0.28}_{-0.23}$ & $14.68^{+0.52}_{-0.43}$\\

\hline 
\botrule
\end{tabular}
\end{table}

\clearpage

\section{Properties of the population of observed satellite trails}\label{appendix:results_prop} 

Table \ref{tab:trail_table} contains the coordinates, orientation and exposure properties of the identified satellite trails in SPHEREx images, obtained between May 2025 and September 2025.

\begin{deluxetable}{cccc}\label{tab:trail_table}
\tablecaption{}
\tablehead{
   \colhead{Row number} & \colhead{Units} & \colhead{Label} & \colhead{Description}
}
\startdata
1 & \nodata & ID & Source identifier \\
2 & deg & ra & Right Ascension in decimal degrees  of the identified trail (J2000) \\
3 & deg & dec & Declination in decimal degrees  of the identified trail (J2000) \\
4 & deg & pa & Position angle of the trail in equatorial coordinates (degrees, PA = 0 is North, counter-clockwise). \\
5 & str & OBS\_ID & SPHEREx Observation ID.\\
6 & -- & MJDOBS & Modified Julian Date of the observation\\
7 & s & XPOSURE & Exposure (integration) time.\\
8 & deg & SGT\_LAT\_MIDPT & Ground track latitude at midpoint.\\
9 & deg & SGT\_LON\_MIDPT & Ground track longitude at midpoint.\\
10 & deg & SGT\_MIN\_GCD2SAA & Closest ground track dist to South Atlantic Anomaly.\\
11 & deg & SPS\_ELON & Ecliptic longitude of planned telescope boresight.\\
12 & deg & SPS\_ELAT & Ecliptic latitude of planned telescope boresight.\\
13 & deg & SGT\_EPA & Position angle of focal plane Y-axis at planned boresight, PA = 0 is North, counter-clockwise.\\
\enddata
\tablecomments{Table \ref{tab:trail_table} is published in its entirety in the electronic edition of the {\it Astronomical Journal}.  A portion is shown here for guidance regarding its form and content. This table is also accessible as a permanent repository at Zenodo: \doi{10.5281/zenodo.17957747}}
\end{deluxetable}


\bibliography{bibFile}{}
\bibliographystyle{aasjournalv7}



\end{document}

%% file: variables.tex
\newcommand{\n}{$-$}

\definecolor{navyblue}{rgb}{0.0, 0.0, 0.5}

\newcommand{\magarc}{mag arcsec\ensuremath{^{\mathrm{-2}}}}

\def\dataset{\def\doi##1{https://doi.org/##1}
\@ifnextchar[{\ydataset}{\xdataset}}
\def\xdataset#1{\ydataset[]{#1}\let\doi\savedoi}
\def\ydataset[#1]#2{\def\one{#1}\ifx\one\empty
\href{#2}{[DATASET]}\else
\href{#2}{#1}\fi\let\doi\savedoi}